\documentclass[a4paper,fleqn,usenatbib]{mnras}
\usepackage[dvipsnames,usenames]{color}

\usepackage[T1]{fontenc}
\usepackage{ae,aecompl}

\usepackage{graphicx}	
\usepackage{amsmath}	
\usepackage{txfonts}    

\defcitealias{HK13}{Paper 1}
\defcitealias{HK14}{Paper 2}
\defcitealias{EHK16}{Paper 3}

\title[Numerical modelling of radio remnants]{Numerical modelling of the lobes of radio galaxies in cluster environments -- IV. Remnant radio galaxies}

\author[W. English et al.]{
W. English$^1$, 
M. J. Hardcastle$^1$\thanks{Corresponding author. E-mail: m.j.hardcastle@herts.ac.uk} and 
M. G. H. Krause$^1$
\\
$^1$Centre for Astrophysics Research, School of Physics, Astronomy and Mathematics, University of Hertfordshire, College Lane, Hatfield, AL10 9AB, UK}

\date{Accepted XXX. Received YYY; in original form ZZZ}

\pubyear{2019}

\begin{document}
\label{firstpage}
\pagerange{\pageref{firstpage}--\pageref{lastpage}}
\maketitle

\begin{abstract}
We examine the remnant phase of radio galaxies using three-dimensional
hydrodynamical simulations of relativistic jets propagating through
cluster environments. By switching the jets off once the lobes have
reached a certain length we can study how the energy distribution
between the lobes and shocked intra-cluster medium compares to that of
an active source, as well as calculate synchrotron emission properties
of the remnant sources. We see that as a result of disturbed cluster
gas beginning to settle back into the initial cluster potential,
streams of dense gas are pushed along the jet axis behind the remnant
lobes, causing them to rise out of the cluster faster than they would
due to buoyancy. This leads to increased adiabatic losses and a rapid
dimming. The rapid decay of total flux density and surface brightness
may explain the small number of remnant sources found in samples with
a high flux density limit and may cause analytic models to
overestimate the remnant fraction expected in sensitive surveys such
as those now being carried out with LOFAR.
\end{abstract}




\section{Introduction}
\label{sec:intro}

The basic mechanism by which the lobes of powerful radio galaxies are
powered has been understood for some time: an accreting active
galactic nucleus (AGN) launches bipolar jets of relativistic plasma
into the surrounding intracluster medium (ICM) \citep{BBR84,BP84}. In
standard models \citep{BR74,S74} these jets impact the local ICM and
end in a termination shock, depositing the spent jet material around
the head of the jet where it goes on to inflate an overpressured
cocoon of low-density plasma which will drive a shell of shocked ICM
around the lobes. Eventually the lobes will come into pressure
equilibrium with the local ICM, but will continue to expand while
energy and momentum is constantly supplied by the central AGN \citep{HW00,C04,HK13}. However, the evolution of radio galaxies once the jets are switched off, known as the remnant phase, is less well understood.

A number of radio-loud AGN have been identified that appear to have no
current jet activity \citep[eg.][]{C87,H97,G07}, and estimates of the
time taken for the remnant source to cool radiatively below
observational limits (up to $10^8$ Myr, \citet{C86}) suggest that this
timescale should be roughly similar to the active phase of the AGN,
though this estimate assumes that radiative losses are dominant,
whereas sources that remain overpressured up to their remnant phase
are expected to dim much more quickly due to adiabatic losses
\citep{KC02}. Observationally these remnant objects remain rare, with
only a small percentage of radio sources being classified as remnants.
\citet{G88} find that in a sample of 187 radio galaxies from the B2
and 3C catalogues, only $\approx 3$ per cent are probably remnants. By
selection, samples with high flux density limits have poor sensitivity
to remnants, but LOFAR observations and modelling give upper limits on
the remnant fraction of $\sim 10$ per cent \citep{B17,Mahatma+18}.
Although the numbers of both are small, true remnant sources may be
even rarer than double-double radio galaxies \citep{S00}, those in
which a new active phase creates lobes clearly distinguishable from
those formed in an earlier phase, despite the short timescale it
should take for the new lobes to merge with the old
\citep{CB91,KSR00,KH13,K13,Mahatma+19}. This suggests that the fading
time of remnant sources is very short.

Numerical modelling of radio galaxies has until recently focused on
the active phase -- understandably since, as seen above, they make up
the vast majority of observed sources. Due to the vastly different
spatial scales needed to model radio galaxies in their entirety, from
the launching of the jets to their impact on the cooling of cluster
gas, simulations are forced to focus on three main areas. Large-scale
cosmological simulations including AGN feedback, either explicitly or
via semi-analytic post-processing, are able to reproduce observed mass
and luminosity functions of galaxies \citep{C05,B06}; however, the
jets are not directly modelled as this would be computationally
infeasible, and instead feedback is included by directly giving the
affected particles an amount of energy (usually either as thermal or a
combination of thermal and kinetic) dependent on values such as the
black hole's mass and an observationally determined feedback
efficiency \citep[eg.][]{V14,S15}. At the other end of the scale are
models which simulate the black hole and its immediate surroundings,
out to a few hundred Schwarzschild radii, which can allow the
investigation of the efficiency of the jet launching process as well
as the energy content of the jets \citep[eg.][]{TNM11}. In between
these are those models which inject an already formed and collimated
jet into some form of environment; from a simple uniform distribution
\citep{N82,L89} to a more realistic beta model \citep{BA03,K05}, and
even using an environment derived from simulations of a dynamically
active cluster \citep{H06,BS17}. Studies of the evolution of radio
lobes that have covered the remnant phase are mostly of the last type
\citep[e.g.][]{BA03,Zanni+05,Perucho+11,Perucho+14}
and, to date, have concentrated on the dynamics of the remnant sources
and their effect on the external medium rather than on their
observational properties.

In previous work \citep[hereafter Papers 1, 2 and 3]{HK13,HK14,EHK16}
we investigated the impact of an active jet, powered by an AGN, on the
surrounding ICM by way of numerical modelling using the last of the
approaches described above. \citetalias{HK13} used two-dimensional,
purely hydrodynamical (HD) models of AGN jets in a range of
environments to study the resulting radio lobes. \citetalias{HK14}
extended this to three-dimensional magnetohydrodynamical (MHD) models,
allowing synthetic synchrotron observations to be created and the
observational properties to be studied in varying environments.
\citetalias{EHK16} took this a step further and used relativistic MHD
(RMHD) models to test the effects of jet power and velocity on the
lobe dynamics and energetics. Here we present results from a set of
three-dimensional RMHD models in which the jet is switched off while
the lobes are still evolving in the cluster environment. The
  inclusion of both relativity and magnetic fields in our models
  allows us to simulate observations of remnant radio sources,
to investigate their impact on the external medium and to and
  make predictions for their observability in different radio
  surveys. In Section 2 we will describe how the simulations are set
up and list the parameters used for the jet and cluster. In Section 3
we present and discuss the results from this set of simulations, and
Section 4 summarizes our findings.

\section{Simulation Setup}

We make use of the freely available code \textsc{PLUTO}\footnote{http://plutocode.ph.unito.it}, version 4.2, described by \citet{M07}. The models are performed using the RMHD physics module, the \textsc{hlld} approximate Riemann solver, and second order dimensionally unsplit Runge-Kutta time stepping algorithm with a Courant-Friedrichs-Lewy number of 0.3. We use the Taub-Matthews equation of state which has a variable adiabatic index which varies with temperature, ranging from the low temperature (ideal gas) value of 5/3 to the high temperature (relativistic plasma) value of 4/3. A divergence cleaning algorithm is used to enforce $\nabla\cdot \mathbf{B}=0$. We do not use the adaptive mesh refinement (AMR) capability of \textsc{PLUTO}.

For consistency with our previous work we define simulation units of
length, $L_0$, and density, $\rho_0$, to be $2.1$ kpc and $3.011
\times 10^{-23}$ kg m$^{-3}$ (giving a unit number density, $n_0$, of
$3 \times 10^4$ m$^{-3}$, for a mean particle mass $\mu$ of $0.6m_p$);
the simulation units for velocity must be the speed of light, $c$, for
the RMHD module. From these the remaining units can be derived, giving
simulation units for time ($t_0 = L_0/v_0$) to be $6.85$ kyr, magnetic
field strength ($B_0 = v_0\sqrt{4\pi \rho_0}$) to be $1.84$ $\mu$T,
and pressure ($p_0 = \rho_0 v_0^2$) to be $2.7 \times 10^{-6}$ Pa. As
in \citetalias{EHK16}, we set the central pressure in the cluster
$p_0$ to be $10^{-10}$ Pa in physical units, or
$(2.7 \times 10^4)^{-1}$ in simulation units.

We simulate a 400 by 400 by 400 element volume ranging between $\pm
150 L_0$, giving a physical resolution of $1.6$ kpc and allowing each
lobe to expand to a length of $315$ kpc. The grid has periodic outer
boundary conditions (in order to enforce zero velocity at the
  boundary), and a cylindrical internal boundary condition is defined
at the centre of the grid, aligned with the $x$-axis, from which the
jet is injected. This internal region has radius $r_j = 2L_0$, giving
a jet resolution of 2.7 cells per jet radius, and length $l_j = 3L_0$.
While this gives an unphysically large jet radius ($r_j = 4.2$ kpc) we
are limited by the numerical resolution, as the jet has to have a high
enough resolution for the internal boundary to couple well to the
local environment. We are more concerned about the physics of the
lobes than those of the jet, and we found, in \citetalias{HK14} and
\citetalias{EHK16}, that these models with a relatively low resolution
across the jet were sufficient to capture the lobe physics well, as
demonstrated by the agreement between \citetalias{HK14} and
\citetalias{EHK16} and other, higher-resolution, models for properties
such as the evolution of the lobes' volume and axial ratio, so in
general we are not concerned by this feature of the models. It is,
however, worth noting that the low resolution across the jet could
lead to numerical diffusion, preventing turbulence and instabilities
which are key in amplifying the injected magnetic field.

Inside the injection region we define the density
$\rho_j$, temperature $T_j$, velocity $\mathbf{v}_j = \pm\mathbf{i} v_x$
($v_y = v_z = 0$) and magnetic field strength
$\mathbf{B}_j$ of the injected material such that the jet is light
compared to the environment, using the values shown in Table \ref{tab:Sim}. A
conserved tracer quantity is injected with the jet material, and
initially has a value of 1.0 inside the injection region and 0
everywhere else. This is used in post-processing to distinguish
between the lobes and shocked cluster material, using the same process
as in \citetalias{EHK16}. The shocked ICM is identified by searching
inwards from each edge of the grid for the surface at which the radial
velocity is greater than $2.5 \times 10^{-4} c$. Similarly, for the
lobes, we search inwards for the surface at which the tracer is
greater than $10^{-3}$, though in the current paper we also search along the jet axis from the centre outwards, with the same criteria, to ensure that any ICM pulled up behind the lobes is not counted as lobe material.

Since we set the internal boundary flag to true inside the injection
region, these cells are not evolved as part of the hydrodynamics
solver. This results in the internal boundary region acting as a sink
into which material and energy can be lost. While the jet is active
this is not an issue, since the material injected by the jet forces the
ICM away from the cluster centre. However once the jet has been turned
off, or even once the lobes have started to rise out from the core,
the ICM will start to relax back into the potential, falling in
towards the internal boundary. To prevent this, once the lobes have
reached the desired length the jets are turned off by disabling the
internal boundary condition and the resulting hole is filled with
stationary material that has density and pressure equal to the average
values found around the outside of the injection region.

The surrounding environment used for these models is given by an isothermal beta model, representing that of a rich group or cluster, with density profile:

\begin{equation}
n=n_0\left[1+\left(\frac{r}{r_c}\right)^2\right]^{-\frac{3\beta}{2}}
\end{equation}

\noindent where $r_c$ is the core radius. In order to break the symmetry between the two lobes we introduce small random perturbations in the density, so that for each set of input parameters we can get results from two non-identical lobes. To keep the cluster environment stable a gravitational acceleration vector is defined by:

\begin{equation}
\mathbf{g}=-\frac{3\beta}{\Gamma\times2.7\times10^4}\frac{\mathbf{r}/|\mathbf{r}|}{\sqrt{\mathbf{r}^2+r_c^2}}
\end{equation}

\noindent where the factor of $2.7\times10^4$ comes from the
temperature of the cluster in simulation units. The magnetic field in
the cluster is set as a Gaussian random field that has an energy
density that scales with thermal pressure, described by \citet{M04}
and \citet{H13}. This is done by generating the Fourier transform of
the magnetic vector potential $\mathbf{A}(k)$. For each of the three
components we draw the complex phase from a uniform distribution and
the magnitude from a Rayleigh distribution whose controlling parameter
depends on $k$. From this we can calculate the Fourier transform of the magnetic field, and by taking the inverse Fourier transform we obtain a divergence-free magnetic field with, after scaling to physical units, a peak field strength at the centre of the cluster of $0.7$ nT.

Two sets of simulations are performed in order to study the behaviour of the lobes once they are no longer being fed by the jet. Simulation parameters for each model are presented in Table~\ref{tab:Sim}.

\begin{enumerate}
\item The main set consists of 9 simulations in which we vary the
  $\beta$ and $r_c$ parameters of the environment in the ranges $\beta
  = 0.55, 0.75, 0.95$ and $r_c = 40, 60, 80$ kpc. The jets in each of
  these models are identical, with the same parameters as the
  v95-med-m model from \citetalias{EHK16}; they are powerful ($Q = 2
  \times 10^{39}$ W), highly relativistic ($v_{jet} = 0.95$, $\gamma_j
  = 3.2$), under-dense ($\eta = 3.478 \times 10^{-6}$) and include a
  toroidal magnetic field around the $x$-axis, where $B_j = 0.209$ nT,
  with $B_y=B_j(z/r)$ and $B_z=B_j(y/r)$ for $r<r_j$. Here $\eta$
    is the relativistic generalization of the jet density contrast
    with the central density of the ambient medium $\rho_0$, as
    described in \citetalias{EHK16}.
    We allow the models to evolve until the lobes have reached a length of $150$ kpc, at which time the jet is shut off as described above. This makes sure that the lobes have sufficient time to form while the jet is active, while ensuring enough time to study the remnant source before it reaches the edge of the simulated grid, at which time the models were terminated.
\item In addition, we run an additional pair of simulations, with the
  same cluster and jet properties as the r75-60 model, but in which
  the jets are turned off once they reach 110 and 190 kpc. This allows
  us to test how the position of the lobes within the cluster affects
  the time taken for the remnant to cool below an observable limit.
  Lobes that are overpressured compared to the local environment are
  expected to dim faster than those closer to pressure equilibrium,
  due to the increased adiabatic expansion losses.
\item To cover the full range of input parameters a final pair of simulations in which the jet power is varied. The first takes the same jet parameters as the v95-low-m model of \citetalias{EHK16}; with jet power $Q = 1 \times 10^{39}$ W, velocity $v_{jet} = 0.95$, density $\eta = 1.739 \times 10^{-6}$ and magnetic field strength $B_j = 0.158$ nT. The second takes the parameters of v95-high-m; jet power $Q = 5 \times 10^{39}$ W, velocity $v_{jet} = 0.95$, density $\eta = 8.696 \times 10^{-6}$ and magnetic field strength $B_j = 0.331$ nT.
\end{enumerate}

\begin{table}
\begin{center}
\caption{Key simulation parameters for each model discussed in this paper. From left to right the columns give the code used to identify each model, the injected jet power $Q$, $\beta$ value and core radius $r_c$ of the cluster environment, and the length $L_{switch-off}$ the lobes are allowed to reach before the jet is switched off.}
\begin{tabular}{ ccccc }
\hline
Code & $Q$ & $\beta$ & $r_c$ & $L_{switch-off}$\\
& (W) & & (kpc) & (kpc)\\
\hline
r55-40 & $2\times10^{39}$ & 0.55 & 40 & 150\\
r55-60 & & & 60 & \\
r55-80 & & & 80 & \\
r75-40 & & 0.75 & 40 & \\
r75-60 & & & 60 & \\
r75-80 & & & 80 & \\
r95-40 & & 0.95 & 40 & \\
r95-60 & & & 60 & \\
r95-80 & & & 80 & \\
r75-60-Early & & 0.75 & 60 & 110\\
r75-60-Late & & & & 190\\
r75-60-Low & $1\times10^{39}$ & & & 150\\
r75-60-High & $5\times10^{39}$ & & & \\
\hline
\end{tabular}
\label{tab:Sim}
\end{center}
\end{table}

As in our earlier work, all of these models were run on the University
of Hertfordshire high-performance computing
facility\footnote{\url{https://uhhpc.herts.ac.uk/}}. 192 Xeon-based
cores were used for each job with the Message Passing Interface (MPI)
being implemented in \textsc{MVAPICH2}. Each model took around 1 week
to run. Calculations are performed in post-processing from output
files produced every 150 simulation time units, or every 1.02 Myr,
consisting of density, velocity, pressure, magnetic field strength and
tracer quantity values for the entire simulation grid. From these the
energetics and dynamics of the lobes and shocked ICM are determined.
We calculate values for each lobe independently due to the slight
density fluctuations in the environment. Therefore lobe properties
presented will be the average value for the two lobes.

\section{Results and discussion}

From here, unless otherwise stated, results will be in physical units. When discussing the dynamics of the lobes we will give results as the average value across the two lobes in each model, but for the observational properties results will be for the entire source. Where we plot results for a single model r75-60 will be used, as this is the most typical representative for the suite of models.

\subsection{Lobe dynamics and energetics}
\label{sec:dynamics}

\begin{figure*}
\includegraphics[width=185mm]{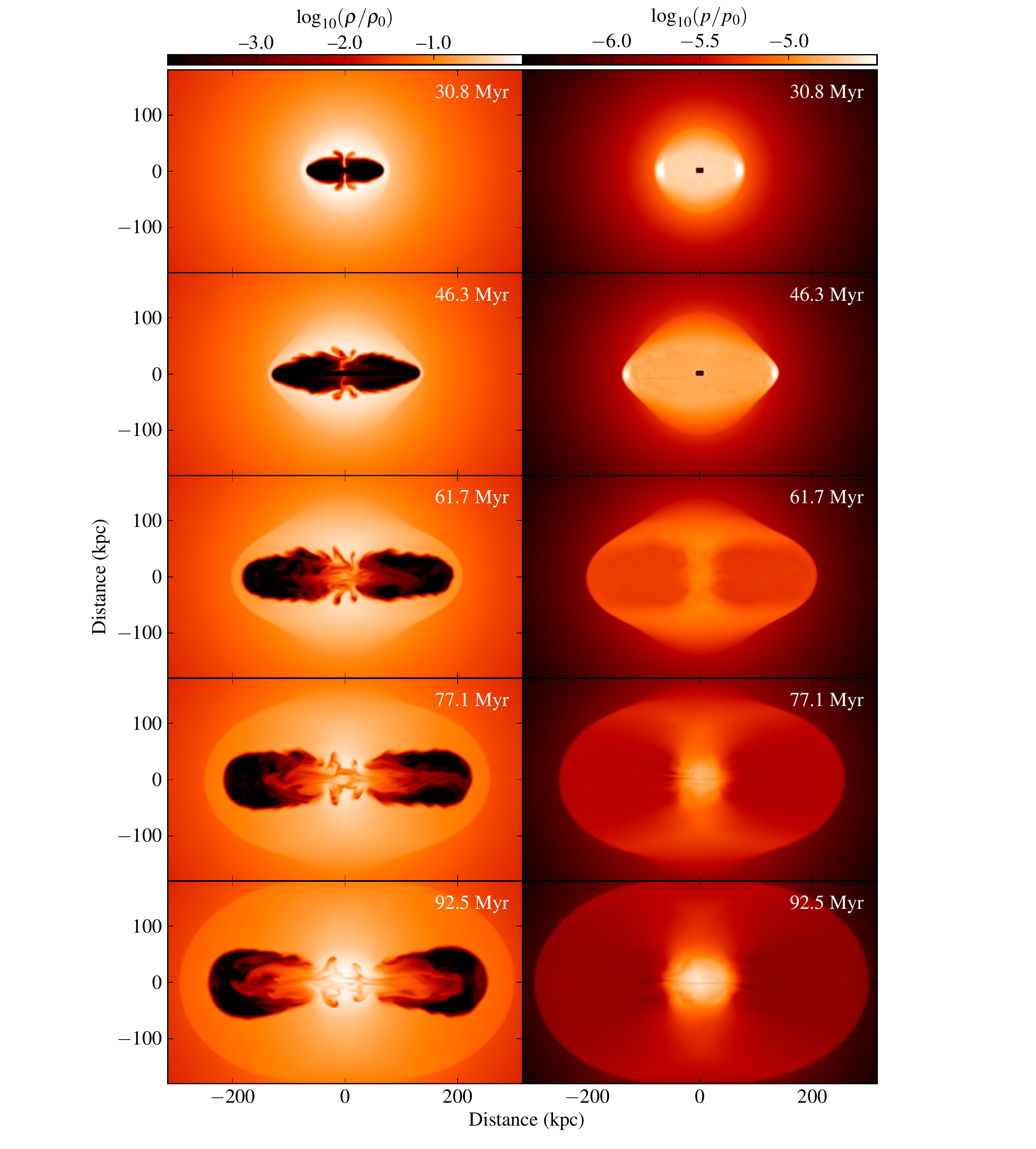}
\caption{Midplane density and thermal pressure slices for the r75-60
  model at 5 different timesteps. Colour scales are logarithmic in
  simulation units of density and pressure, ranging from $-4.0$ to
  $0.0$ in density and from $-6.5$ to $-4.5$ in pressure. The jet is
  switched off once the lobe reaches $150$ kpc, which in this model
  occurs at $50$ Myr. Timestamps give the age of the model at each step. Not included in these plots is the contribution to the pressure from the magnetic field, which accounts for the slight underpressure of the lobes that appears in these plots.}
\label{fig:Dens}
\end{figure*}

\begin{figure*}
\includegraphics[width=170mm]{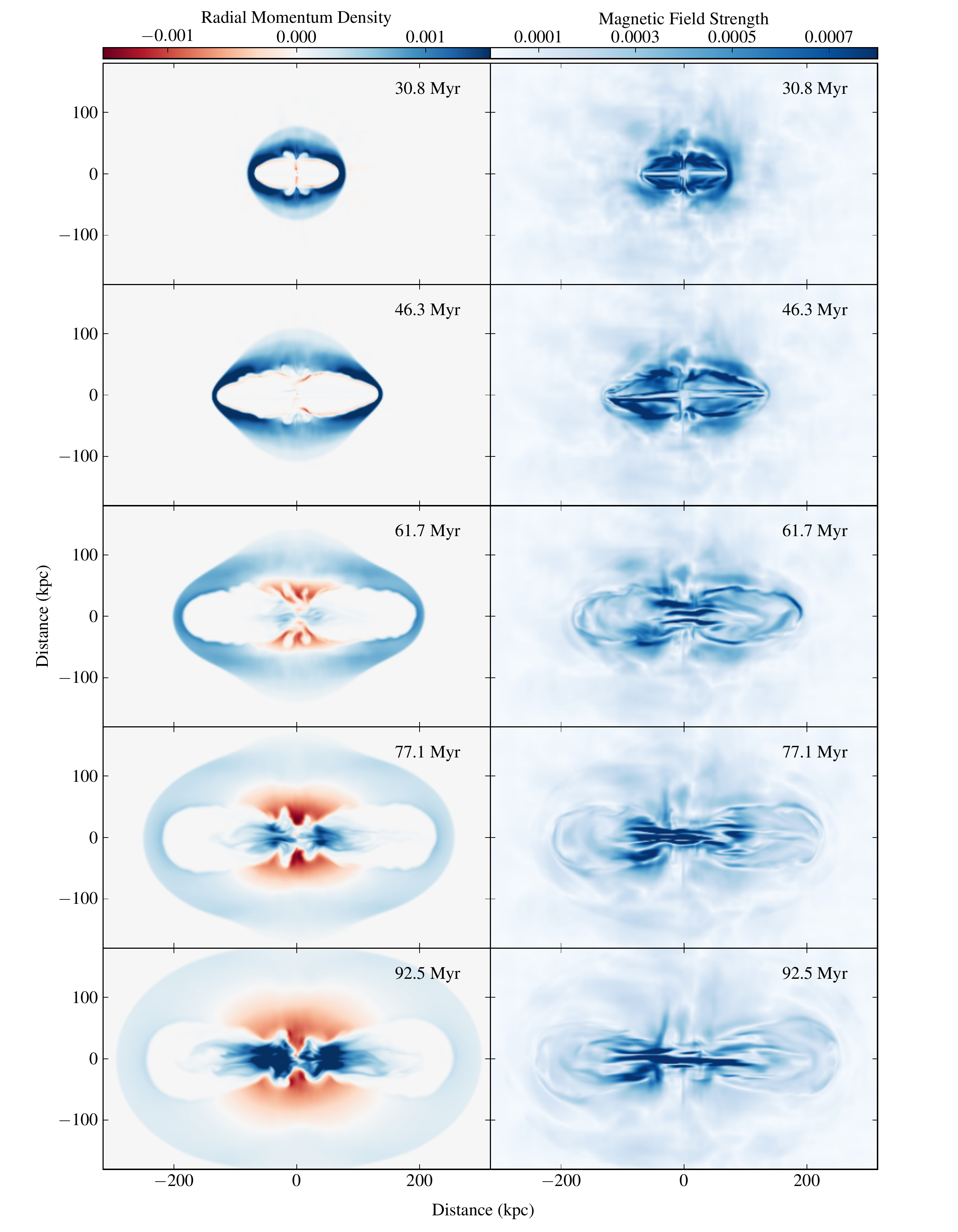}
\caption{Midplane magnitude of the radial momentum density and magnetic field strength slices for the r75-60 model, at the same timesteps as Figure~\ref{fig:Dens}. Colour scales are linear in momentum and magnetic field strength, both in simulation units, ranging from $-0.0015$ to $0.0015$ in momentum density and from $0.0$ to $0.0008$ in magnetic field strength.}
\label{fig:Momentum}
\end{figure*}

\begin{figure*}
\includegraphics[width=175mm]{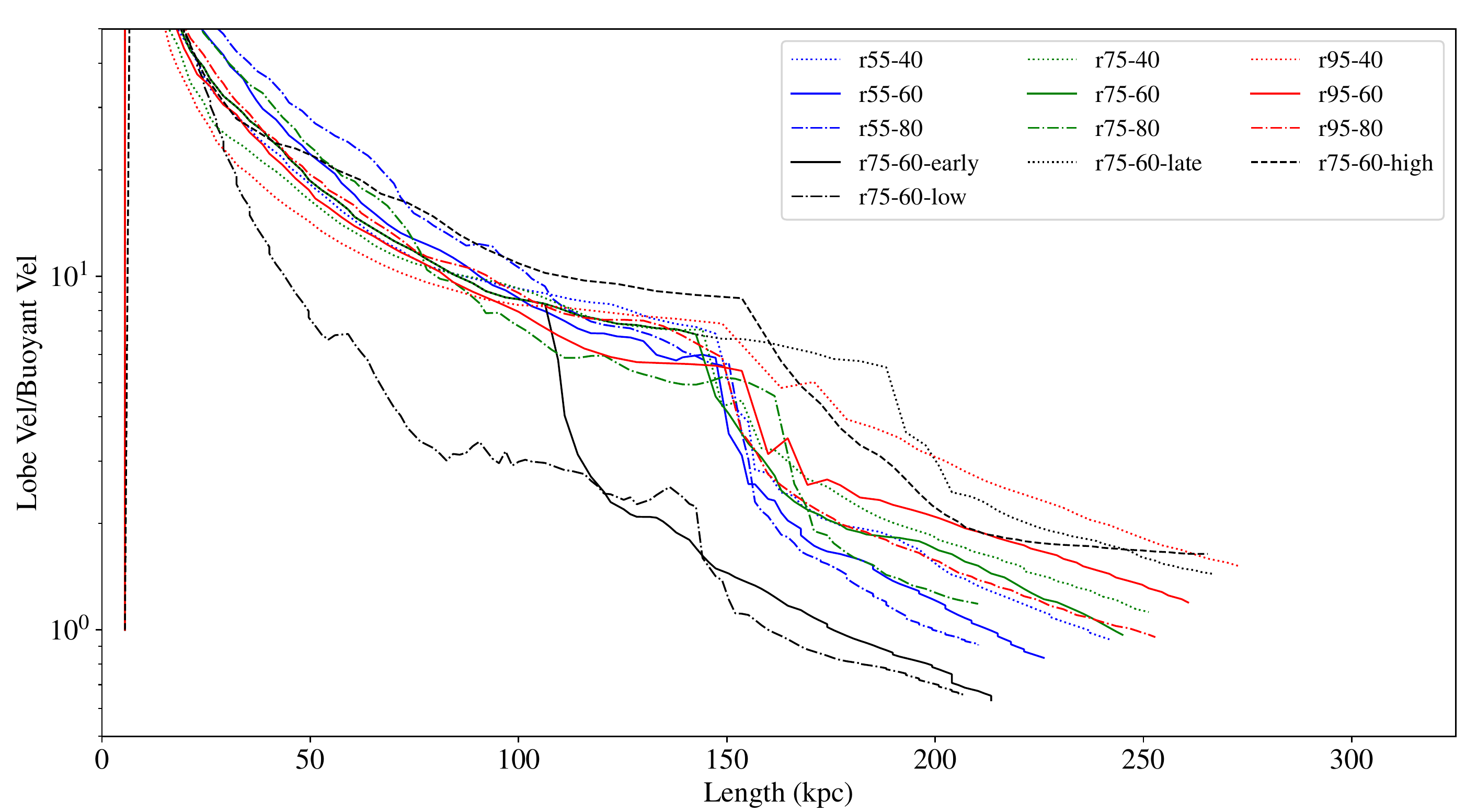}
\caption{Ratio of the advance velocity of the lobes in each model to the velocity of a bubble of low density gas, at the same position within the cluster as the lobes, moving purely due to buoyancy showing that for a significant amount of time after the jets have been switched off the lobes continue to be forced out of the cluster by more than just buoyancy.}
\label{fig:Buoy}
\end{figure*}

We begin by discussing the lobe dynamics during both the active and
remnant phases. Figure~\ref{fig:Dens} shows snapshot midplane density
and pressure slices for a number of timesteps for the r75-60 model,
showing the evolution of the lobes as they are inflated by the jet and
later buoyantly rise out of the centre of the cluster. The black
square visible in the first two pressure maps is the injection region,
which is turned off and filled in once the lobes reach a length of
$150$ kpc for the main suite of models, which occurs at $50$ Myrs
here. We can see that while the jet is active the lobes are visible
against the surrounding cocoon of shocked ICM in the pressure maps, as
during this phase they are still overpressured with respect to their
environment. Once the jets are switched off they very quickly reach
pressure equilibrium. Switching off the jet also has a large effect on
the shape of the shock driven through the cluster; as there is no
longer a much stronger force pushing the shock in the longitudinal
direction, it quickly takes on a more spherical shape. In the density
maps we see that as the unpowered lobes are floating outwards they do
not keep their prolate spheroidal shape, but instead take on more of a
`U' shape as some of the dense cluster material being dragged along by
the jet pushes inside the back of the remnant lobe and hollows it out.
This can be better seen in the left side of Figure~\ref{fig:Momentum},
which shows midplane slices of momentum density for the same timesteps
for the r75-60 model. Here we clearly see in the last two plots that a
large amount of dense cluster gas is falling back into the potential
and refilling the cluster's initial density distribution, but the
weight of this infalling gas is also pushing `streams' of material
along the channel left by the jets and into the lobes. For some models
this dense gas has pushed all the way through to the front of the
lobes, breaking them apart, by the end of the simulations.
Combined with the fact that the lobes are still overpressured at
  their tips after the switch-off of the jet, this has
the effect of pushing the lobes out faster than they would buoyantly
rise (as demonstrated in Figure~\ref{fig:Buoy}), driving them out to lower
pressure regions of the cluster where they will undergo faster
adiabatic expansion, and in some
cases destroying the lobes before they reach the edge of the grid.

The right hand side of Figure~\ref{fig:Momentum} shows the evolution
of the magnetic field strength. While the jets are active the majority
of the magnetic energy density is located in the lobes, as they are
constantly being fed by the injected toroidal field. Also visible is
the very strong field that forms around the edge of the lobes, caused
by the compression of the initial cluster field by the shock. After a
short while in the remnant phase the magnetic energy density in the lobes drops significantly, due to the expansion of the lobes with no new energy being added. Instead the majority of the magnetic energy now comes from the compressed and shear-amplified cluster field which has been pulled around the back of the lobes and is settling back in the centre of the cluster.

\begin{figure*}
\includegraphics[width=175mm]{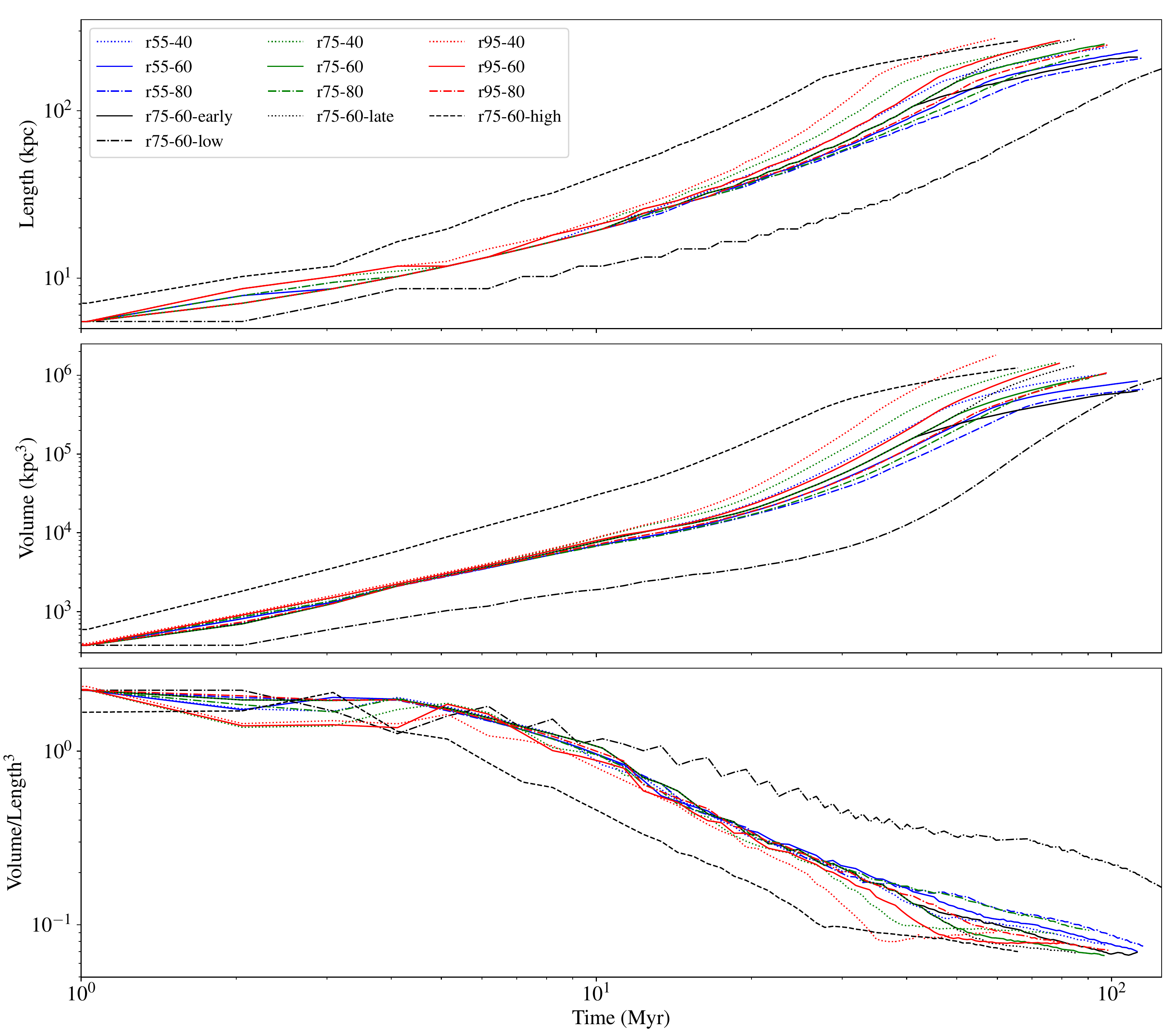}
\caption{Growth of the lobes with time for all of the simulations run in this paper, in terms of the lobes' length (top), volume (middle) and axial ratio (bottom). For each model the plotted value is the average value for the two lobes.}
\label{fig:RHDdyn}
\end{figure*}

The dynamics of the lobes (Fig.\ \ref{fig:RHDdyn}) are, as expected, the same as in
\citetalias{HK14} and \citetalias{EHK16} during the active phase for
the different environments. In models with a smaller core radius the
lobes grow faster both in terms of length and volume, as they quickly
leave the dense core and push out into the lower-pressure outer
cluster. Similarly, environments with a steeper density gradient
(higher values of $\beta$) lead to faster lobe growth at late times. Once the jets are turned off the lobes quickly slow their growth, with environment having little effect on the growth of the lobes: all of the models settle into the same expansion speed.

\begin{figure*}
\includegraphics[width=175mm]{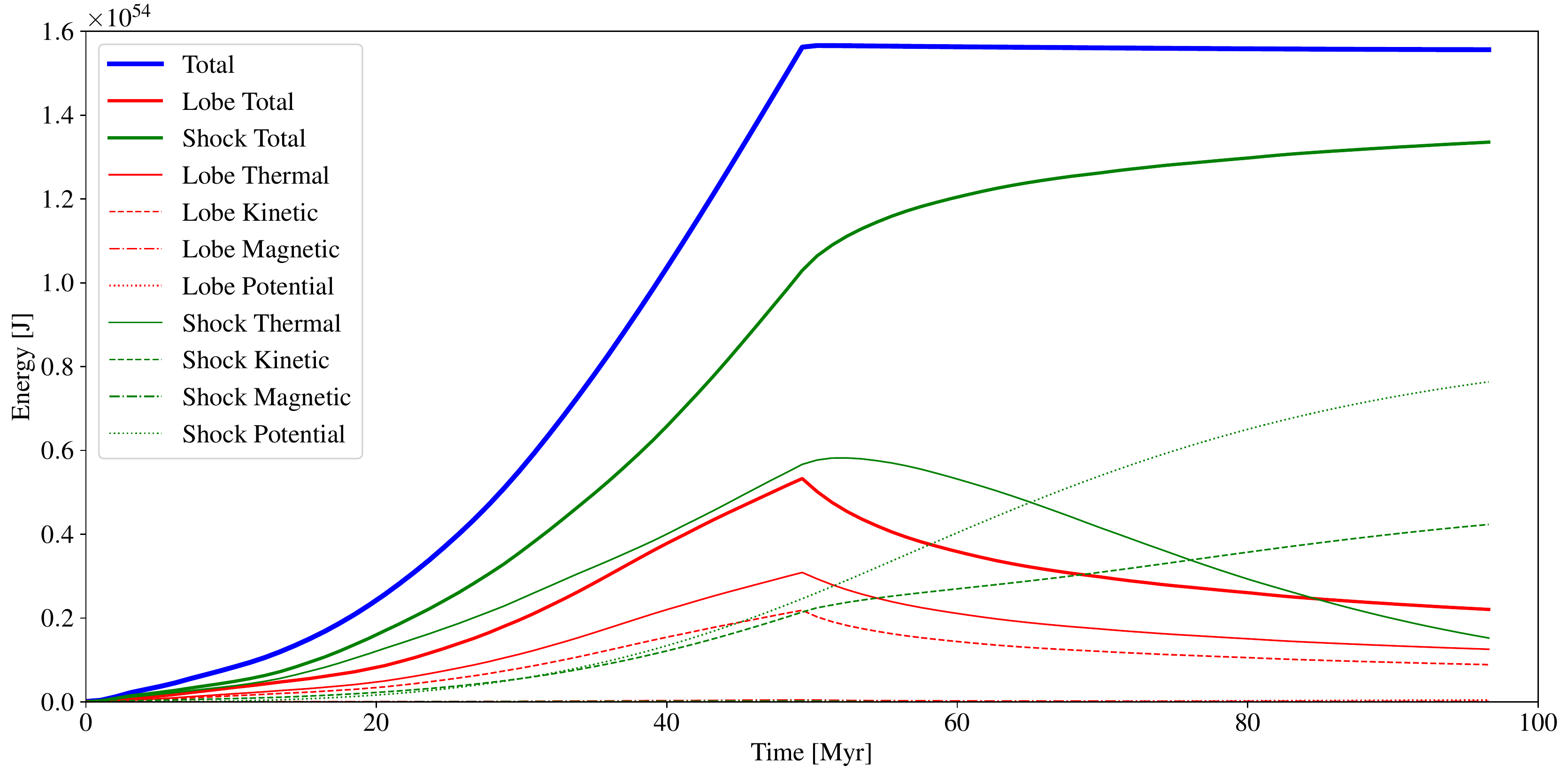}
\caption{Evolution of the energy stored in the lobes and shocked
  cluster regions for the r75-60 model. As in previous models the
  internal boundary does not couple well with the external environment
  for some time, resulting in less than the intended injected energy
  making it onto the grid until the lobes are reasonably well formed.
  Once the jet is switched off, at $t \approx 50$ Myr, energy is lost from the
lobes and gained by the shocked medium.}
\label{fig:Energy}
\end{figure*}

\begin{figure*}
\includegraphics[width=165mm]{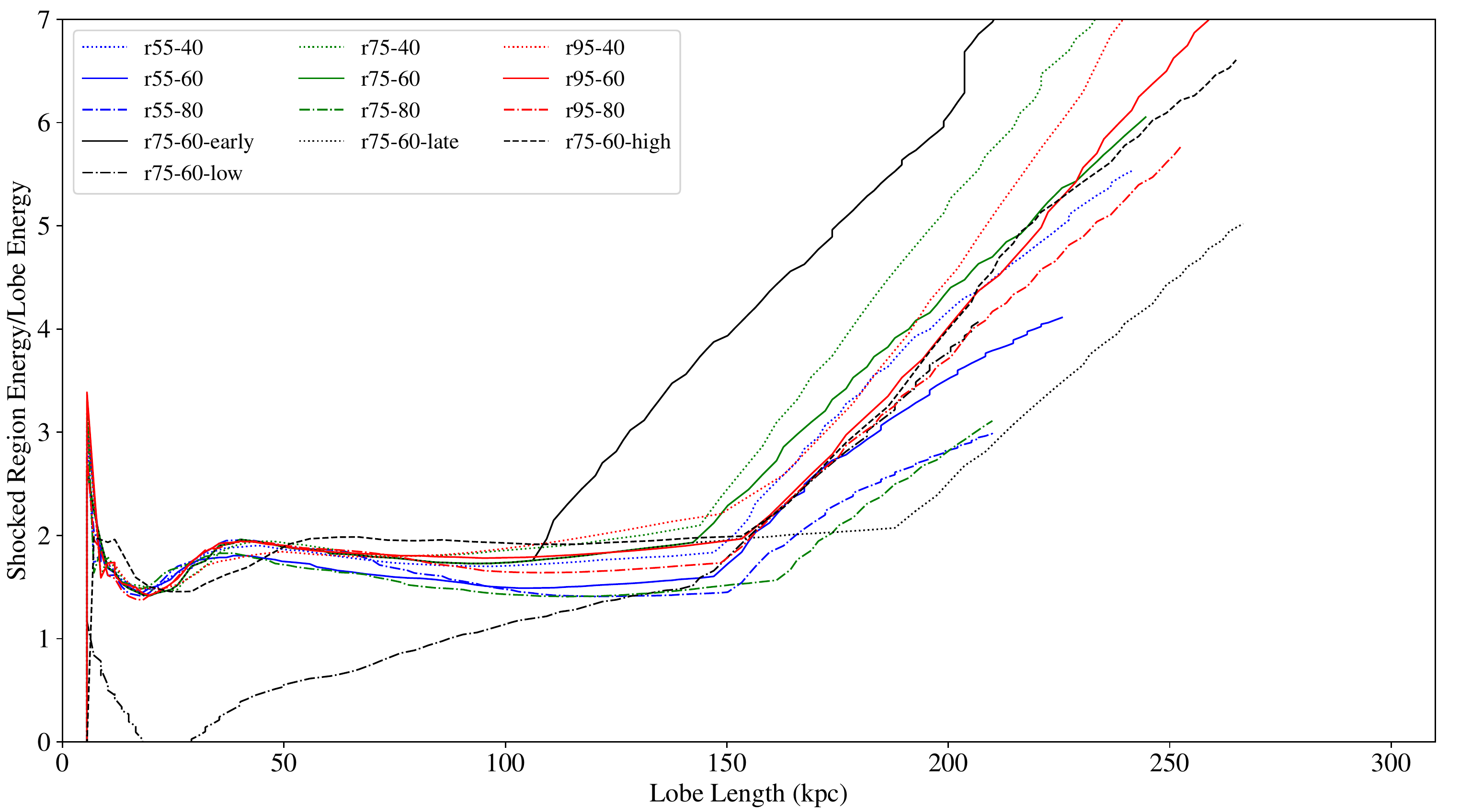}
\caption{Ratio of energy stored in the shocked ICM to that stored in
  the lobes for all of the models, as a function of lobe length. The
  sharp increase in the ratio (at lengths $> 150$ kpc in most models) takes place after the jet is switched off.}
\label{fig:ShockLobe}
\end{figure*}

Figure~\ref{fig:Energy} shows the evolution of the different forms of
energy with time for the r75-60 model. Due to difficulty getting the
internal boundary and the external environment to couple well at early
times, the total injected energy is lower than the expected value
until the lobes have formed and the shock has evacuated the material
immediately surrounding the injection region. Since the fraction of
energy being injected in each of the different components stays
constant during this time, and since the focus of this paper is on the
remnant phase, this early expansion should have little effect on our
results. The cluster's energy content, however, undergoes a
significant change during the remnant phase. As soon as the jet is no
longer actively driving the shock the temperature of this thermal gas
drops and it begins falling back down the cluster potential. We can also see
that immediately after the jets switch off the total energy stored in
the lobes drops significantly, as they continue to expand and do work
on the ICM but without their own supply of energy. As the lobes approach
pressure equilibrium with the cluster gas, however, the rate at which
they transfer energy to the ICM slows. This is also seen in
Figure~\ref{fig:ShockLobe}, which plots the ratio of energy stored in
the cluster to that in the lobes. As in the previous models with a
relativistic equation of state we find
that the ratio of energy for an active well-formed lobe is in the
range $1.5$ to $2$, with environments with a smaller core radius and
steeper density gradient leading to a higher percentage of the energy
going into the cluster. Once the jet is switched off the ratio rapidly
increases; interestingly, this occurs at roughly the same rate regardless of
cluster properties or even the lobes' position within the cluster once
the jet is switched off. Part of the reason for this transfer could in
principle be mixing between the lobe and cluster material, and the way
in which we identify the two regions: if sufficient mixing occurs to
drop the value of the tracer in a cell below our threshold the cell
will no longer be flagged as part of the lobe, and that energy will be
counted towards the cluster's total instead. However, we believe that
the transfer of energy is in fact is a genuine consequence of the
altered lobe dynamics in the remnant phase. Overall mixing is seen to
have only a minor effect on the amount of energy stored in each
region, with the calculated ratio being consistently within the range
$1.5$ to $2$ when the value of the tracer threshold is varied. The
bulk of the energetic effect seems to be due to the bulk increase of
the potential and kinetic energy of the shocked material at the
expense of the lobes. Similar effects are seen in other
  simulations of the same dynamical situation \citep{Perucho+11}.

The expansion of remnant radio lobes has a broad similarity to a Sedov
explosion and can be qualitatively understood by considering the
energetics of a thin-shell model \citep{Krause03,Krause+Diehl14}: a
blast wave driven by a time-dependent energy input $E(t) = {\cal
  L}t^d$, where $\cal L$ is some dimensionally appropriate normalizing
factor, in a spherically symmetric environment with density $\rho(r
)\propto r^{\kappa}$ has a kinetic energy fraction of $\epsilon_\mathrm{k}=\frac{d+2}{(d+1)(\kappa+5)}$. The kinetic energy is, in this approximation, in the shell, and so $r_\mathrm{SL}=\epsilon_\mathrm{k}/(1-\epsilon_\mathrm{k})$ can be taken as a proxy for our ratio between energy in shock and lobes. First neglecting gravity, $d=0$ and $\kappa = -2.25$, as appropriate for a remnant in our setup, leads to $r_\mathrm{SL}=2.7$. In our simulations, energy is however continuously lost to gravitational potential energy, which is not taken into account in the standard thin-shell blastwave solution. We can, however, take this energy loss of the blastwave system to gravitational potential energy into account qualitatively, by using a negative $d$. This leads to a further increase of $r_\mathrm{SL}$ in qualitative agreement with our findings.

\subsection{Observational properties}

Using the same process as in \citetalias{EHK16} we calculate the
synchrotron emissivities for the different Stokes parameters for each cell in the simulation grid. Since the emission is anisotropic this must be repeated across a range of viewing angles, achieved by defining a projection vector pointing from the centre of the grid to the observer. The synchrotron emission for a given cell is then calculated for a transformed version of this projection vector to take into account relativistic aberration, where the apparent position of an observer in the reference frame of an object moving at relativistic speeds differs to the position of the observer in the lab frame. From this we calculate the Stokes $I$ (total intensity), $Q$ and $U$ (polarized intensities) parameters (in simulation units) using the following equations:

\begin{equation} \label{eq:SI}
j_I=p(B_x^2+B_y^2)^{\frac{\alpha-1}{2}}(B_x^2+B_y^2)D^{3+\alpha}
\end{equation}

\begin{equation} \label{eq:SQ}
j_Q=\mu p(B_x^2+B_y^2)^{\frac{\alpha-1}{2}}(B_x^2-B_y^2)D^{3+\alpha}
\end{equation}

\begin{equation} \label{eq:SU}
j_U=\mu p(B_x^2+B_y^2)^{\frac{\alpha-1}{2}}(2B_xB_y)D^{3+\alpha}
\end{equation}

where, for a fixed power-law electron energy distribution, the local thermal pressure $p$ is proportional to the electron number density. $\alpha$ is the power-law synchrotron spectral index (taken to be $\alpha = 0.5$) and $\mu$ is the maximum fractional polarization (equal to $\mu = 0.69$ for $\alpha = 0.5$). $D$ is the Doppler factor, given by:

\begin{equation} \label{eq:Doppler}
D=\frac{1}{\gamma(1-\beta\cos(\theta))}
\end{equation}

where $\beta = v/c$ and $\theta$ is the angle between the projection vector and the velocity vector of the cell. This factor is then raised to the power $(3+\alpha)$ when calculating the emissivities to account for the increased rate at which photons are received in the lab frame compared to the rate they are emitted, the boosting of these photons to higher energies and the fact that the emitted radiation is preferentially beamed towards the direction of motion. To convert this into a more useful, physical, unit we again use a modified form of the equation from \citetalias{HK13}:

\begin{equation} \label{eq:SyncPhys}
j_0 = c(q)\frac{e^3}{\epsilon_0 c m_e} \left(\frac{\nu m_e^3 c^4}{e}\right)^{-\frac{q-1}{2}} \frac{3p_0}{4\pi I}\left(\frac{B_0^2}{8\pi\mu_0}\right)^{\frac{q+1}{4}} L_0^3
\end{equation}

where $c(q)$ is a dimensionless constant ($\approx 0.05$), $e$ and
$m_e$ are the charge and mass of an electron, $\epsilon_0$ and $\mu_0$
are the permittivity and permeability of free space and $c$ is the
speed of light. For spectral index $\alpha = 0.5$ the electron energy
power-law index $q$ is $2$, $\nu$ is the frequency at which we observe
the source and $p_0$, $L_0$ and $B_0$ are simulation units of
pressure, length and magnetic field strength, respectively. Finally
$I$ is the integral over $EN(E)$ between $E_{min}$ and $E_{max}$, with
$E_{min} = 10m_ec^2$ and $E_{max}=10^5m_ec^2$. From this we get the
simulation unit of radio luminosity to be $j_0 = 3.133\times10^{31}$ W
Hz$^{-1}$ sr$^{-1}$ at 150 MHz.

\begin{figure}
\includegraphics[width=85mm]{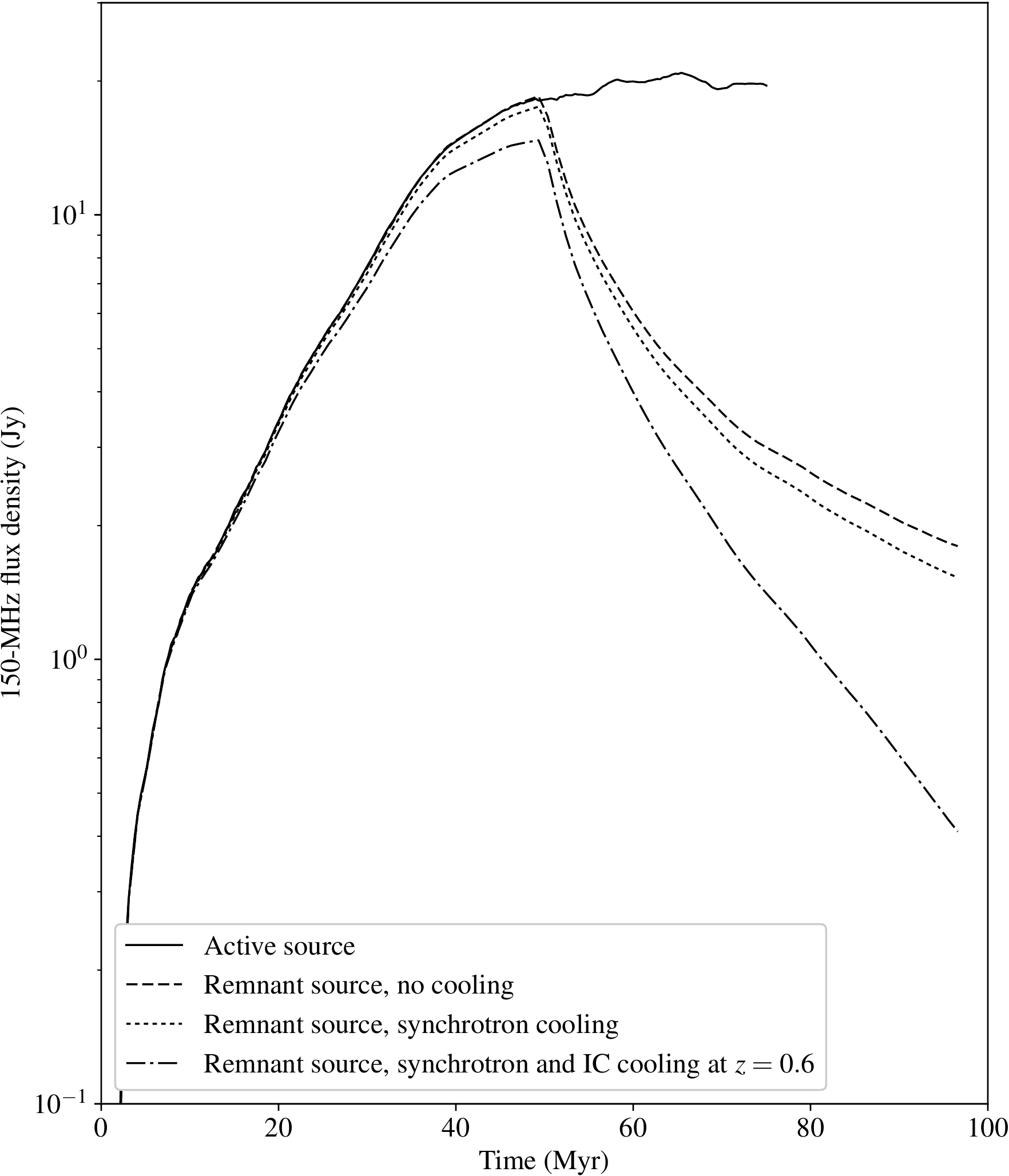}
\caption{Synchrotron flux density evolutions for the v95-med-m model of \citetalias{EHK16} without cooling and the r75-60 model without cooling, with only synchrotron cooling and with both synchrotron and inverse-Compton losses assuming the source is at redshift $z=0.6$.}
\label{fig:Lumi1}
\end{figure}

To allow us to compare these results with observations, such as those
of \citet{H16}, we convert these values into units of Jy. To do this
we assume a standard flat $\Lambda$CDM cosmology, where $H_0 = 70$km
s$^{-1}$, $\Omega_m = 0.3$ and $\Omega_\Lambda = 0.7$. We place the
models at redshift $z = 0.6$, a fairly typical value for the AGN
identified by \citet{H16}. Using this we obtain the simulation unit of
radio flux density, $f_0 = 9.185\times10^5$Jy at 150 MHz. Results are
then compared to the 3CRR flux density limit of 10 Jy and to a surface brightness
limit of 300 $\mu$Jy beam$^{-1}$ for a $6$-arcsec Gaussian restoring
beam, matched to the properties of the LOFAR Two-Metre Sky Survey,
LoTSS \citep{Shimwell+17,Shimwell+19}. We include the effects of
cooling due to synchrotron and inverse-Compton emission using the code
from \citet{H18}, which calculates a time-dependent correction factor
that is applied to our synchrotron emissivities assuming that a
population of electrons in a power-law distribution is supplied at a
constant rate while the jet is active, and uses the average magnetic
field strength history in the lobes. These cooling processes have a
significant effect on the calculated luminosities of our models
(Figure~\ref{fig:Lumi1}); for a model at a redshift of $z=0.6$ the
luminosity has dropped by around an order of magnitude more when
cooling is included, mostly due to the inverse-Compton losses.
However, relative to the analytic models of \cite{H18}, the {\it
  uncorrected} synchrotron luminosity drops more rapidly once the jet
is switched off. We attribute this to the effects of the returning
cluster-centre gas on the dynamics, as discussed in Section
\ref{sec:dynamics}, since the continued pressure-driven expansion of
the remnant lobes is accounted for in the analytical models.

\begin{figure*}
\includegraphics[width=175mm]{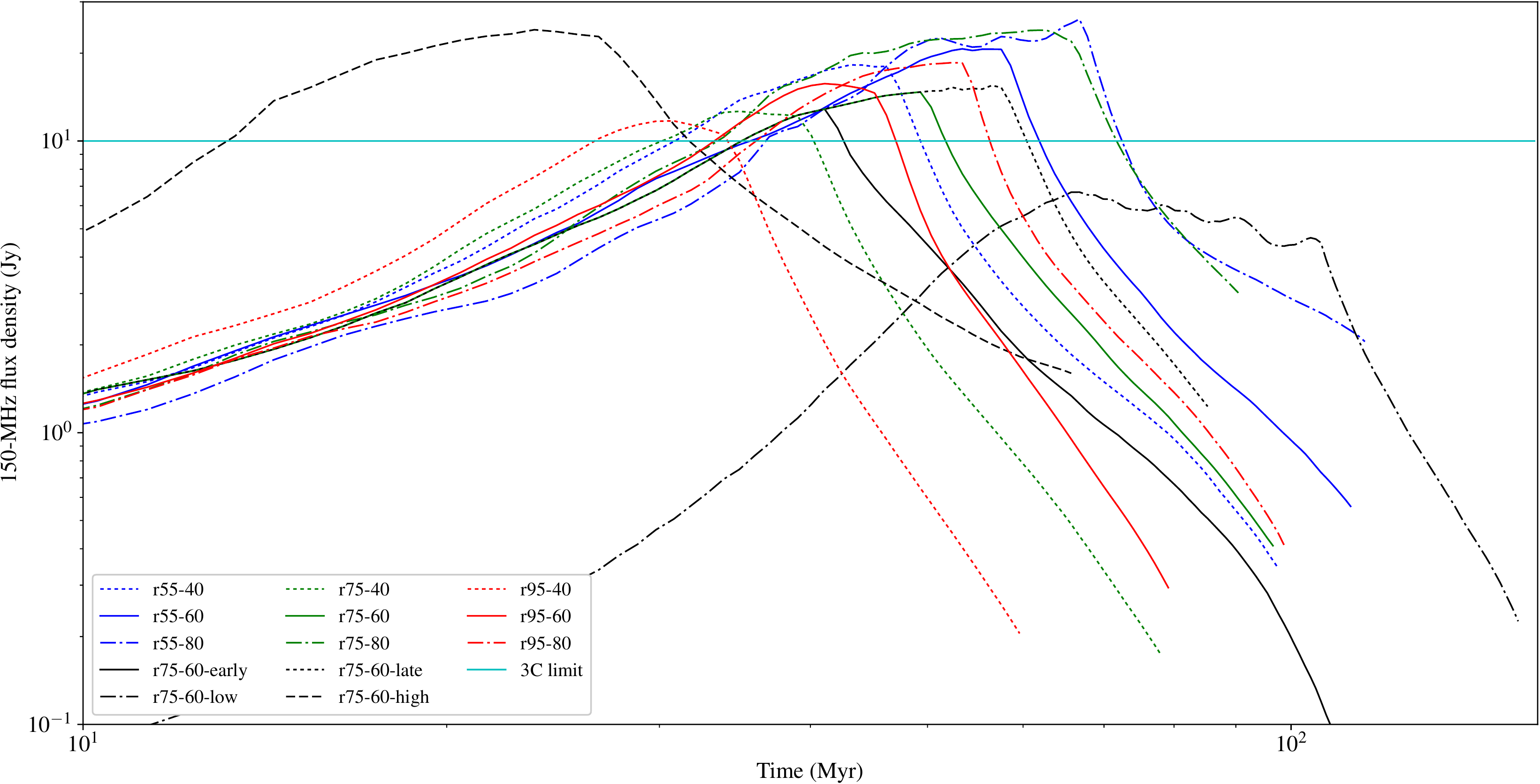}
\caption{Evolution of the synchrotron flux density with time for the entire suite of simulations, assuming they are at $z = 0.6$.}
\label{fig:Luminosity}
\end{figure*}

In Figure~\ref{fig:Luminosity} we show the synchrotron light curves
for each of the models, plotted in terms of rest-frame 150-MHz flux density at $z=0.6$. The expected trends are again
repeated here while the jet is active: environments with a larger core
radius lead to brighter sources while the steepness of the environment
has little effect, with shallower slopes possibly leading to higher
luminosities. Once the jets are switched off the luminosity drops
rapidly for all of the models, falling by around an order of magnitude
by the time the lobes have grown by a further $50$ kpc. The rate at
which the luminosity decays is very similar for each model, with the
main difference being for the models where $\beta=0.55$, which after
the initial drop all start to flatten out more noticeably than the
other models. We see that for these powerful sources in rich
environments, as expected, the total flux density at the peak of
activity would put them in the category of 3C sources (except for r75-60-low). However, within
a few Myr of the cessation of jet activity, their total flux density
drops below the 3C limit. This behaviour may go some way towards
explaining the very low fraction of candidate remnant sources seen in
catalogues selected with a high flux limit (Section \ref{sec:intro}).

\begin{figure*}
\includegraphics[width=175mm]{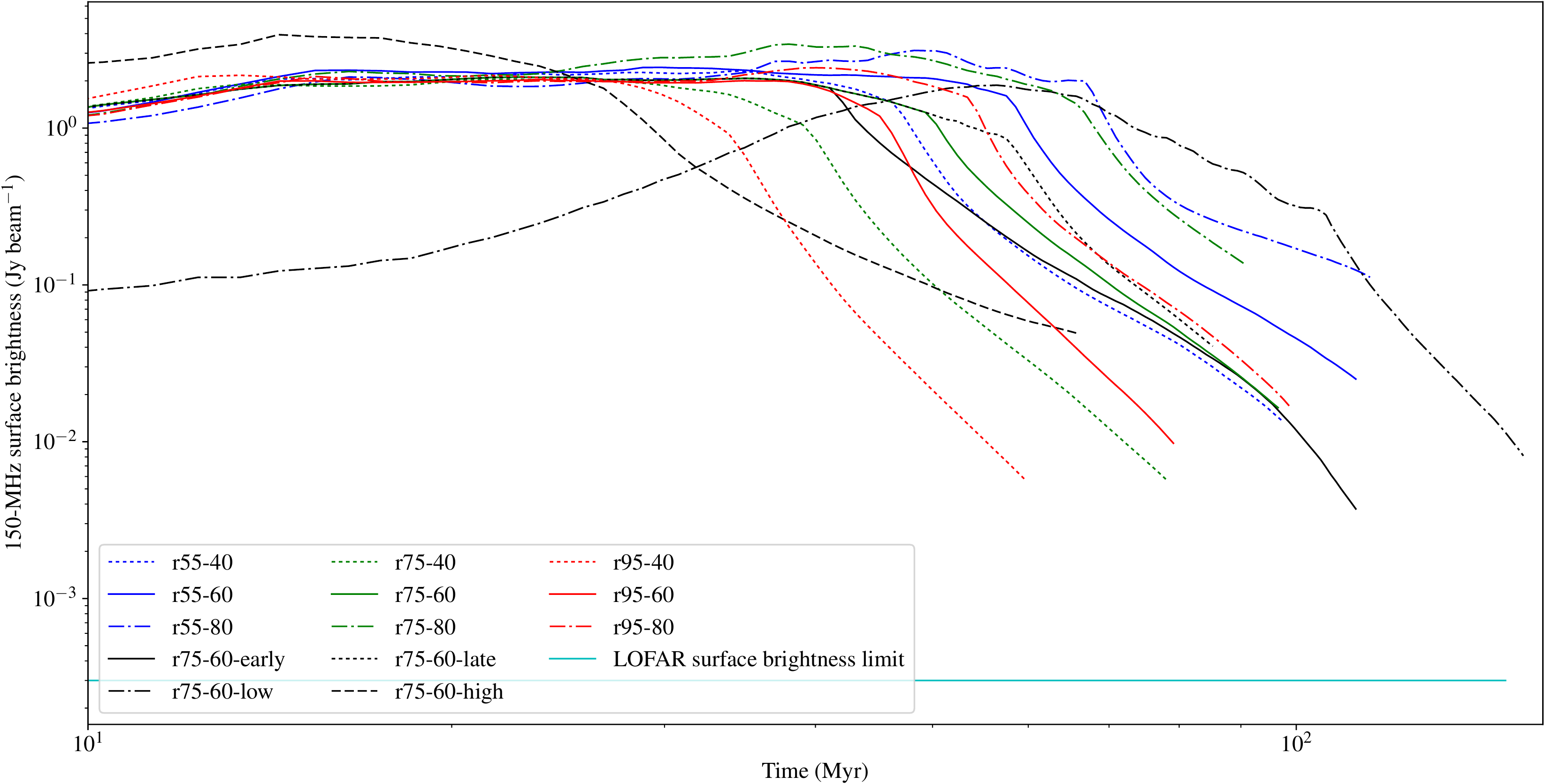}
\caption{Evolution of the mean surface brightness with time for each model, assuming they are at $z = 0.6$ and for a $6$-arcsec Gaussian restoring beam.}
\label{fig:SurfBright}
\end{figure*}

To get a better idea for how quickly the remnant source fades we look
at how the mean surface brightness of each models evolves with time
(Figure~\ref{fig:SurfBright}). From
this plot there is a clear dependence on both environment properties, in
the sense that flatter density profiles and larger core radii imply
lower surface brightness gradients. A
steeper declining density and pressure gradient means that by the time
the jets are switched off, which occurs once the lobes reach a fixed
length of $150$ kpc, the lobes will be contained within lower-pressure
cluster material and so will undergo a faster stage of adiabatic
expansion than those models that are within a denser environment.
During this adiabatic expansion the surface brightness drops rapidly
as both the magnetic field strength and the particle energies decrease
\citep[eg.][]{KC02}. Similarly we see that a large core radius has the
opposite effect. This is essentially for the same reason: the larger
the core radius the closer to pressure equilibrium the lobes will be
once the jets are turned off, therefore the expansion of these lobes
in the remnant phase is expected to be much more relaxed. Further
evidence for this comes from the two models in which the jets were
switched off earlier and later. While we had expected that the earlier
model (r75-60-early) would still be highly overpressured, having had
little time to establish equilibrium with its environment, and would
therefore cool faster, the results are actually the other way around:
the position within the environment appears to have a greater impact
on the amount of time spent in the remnant phase than the stage of the
lobes' evolution does.

Comparing the actual values of surface brightness for sources at
$z=0.6$ to the LOFAR $3\sigma$ surface brightness limit, we see that
powerful sources like these should remain detectable to a sensitive
instrument like LOFAR for a long time. The faintest simulated objects
remain an order of magnitude above the LOFAR limit at the end of the
simulation, after an elapsed time significantly longer than their
active lifetime. Hence, if remnant powerful sources of this type exist, we
would expect to find them in LOFAR observations in numbers which would
be comparable to or even exceeding those of the parent luminous radio
galaxy population. Two caveats are necessary to this statement:
firstly, at significantly higher redshift, the effect of
inverse-Compton losses would become more severe and would suppress the
existence of relics; the second is that, as noted above, the
environment of the source and the timing of the termination of jet
activity significantly affect the fading timescale. In practice it
seems likely that many of these sources will be affected by subsequent
restarts of jet activity, a situation recently modelled by \cite{Yates+18}.

Because we have modelled a relatively narrow range in jet power,
focussing on the high-power end of the range of realistic radio
sources, we cannot directly comment on the remnant fractions seen in
LOFAR data as did \cite{Mahatma+18}, \cite{GMB17} and \citet{H18}. We
have also considered only a narrow range of shutoff times, as a result
of our choice to terminate jet activity when the source reaches 150
kpc. As shown by \cite{Hardcastle+19}, the properties of low-$z$ radio
galaxies are consistent with having a more or less uniform
distribution of active times up to 1 Gyr, and a more realistic
calculation of remnant fraction would require us to simulate jets that
switch off over a much wider distribution of lifetimes and of jet
powers than is practical in these models. Qualitatively, however,
since we find a faster initial drop in the synchrotron luminosity in
our simulations than is seen in analytical models, we might expect the
true remnant fraction at low $z$ to be lower than the $\sim 50$ per
cent of \cite{GMB17} or the $\sim 30$ per cent calculated by
\citet{H18}, and we suggest that a consideration of these effects
would bring the models closer to the observations.

\begin{figure*}
\includegraphics[width=175mm]{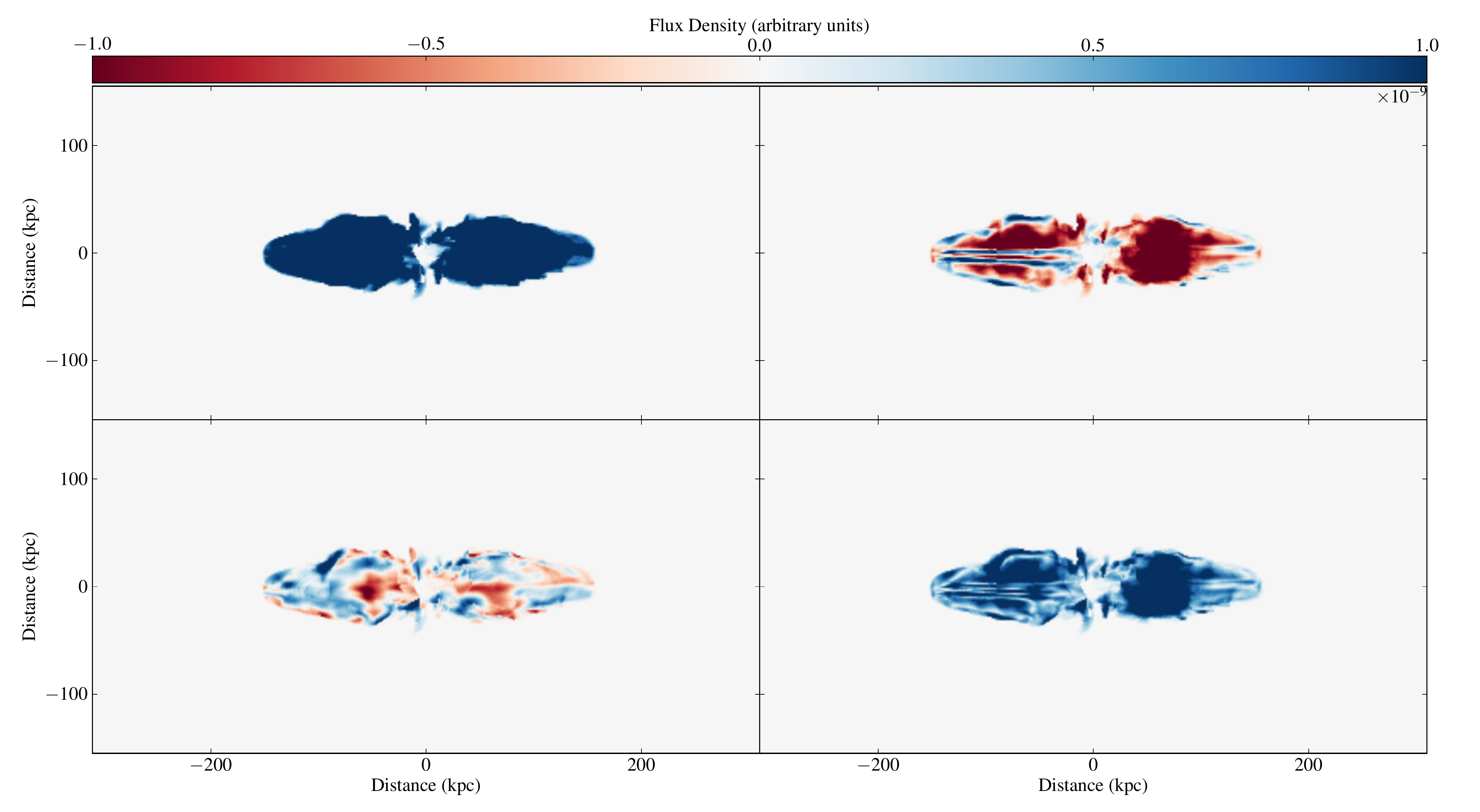}
\caption{Synchrotron emission maps for the r75-60 simulation, observed at 90 degrees to the jet axis at an age of 51.3 Myr, just after the jets have been switched off. Top row: Stokes $I$ (left) and $Q$ (right). Bottom row: Stokes $U$ (left) and $P=\sqrt{Q^2+U^2}$ (right). All maps are scaled by the same arbitrary amount.}
\label{fig:SynchEarly}
\end{figure*}

\begin{figure*}
\includegraphics[width=175mm]{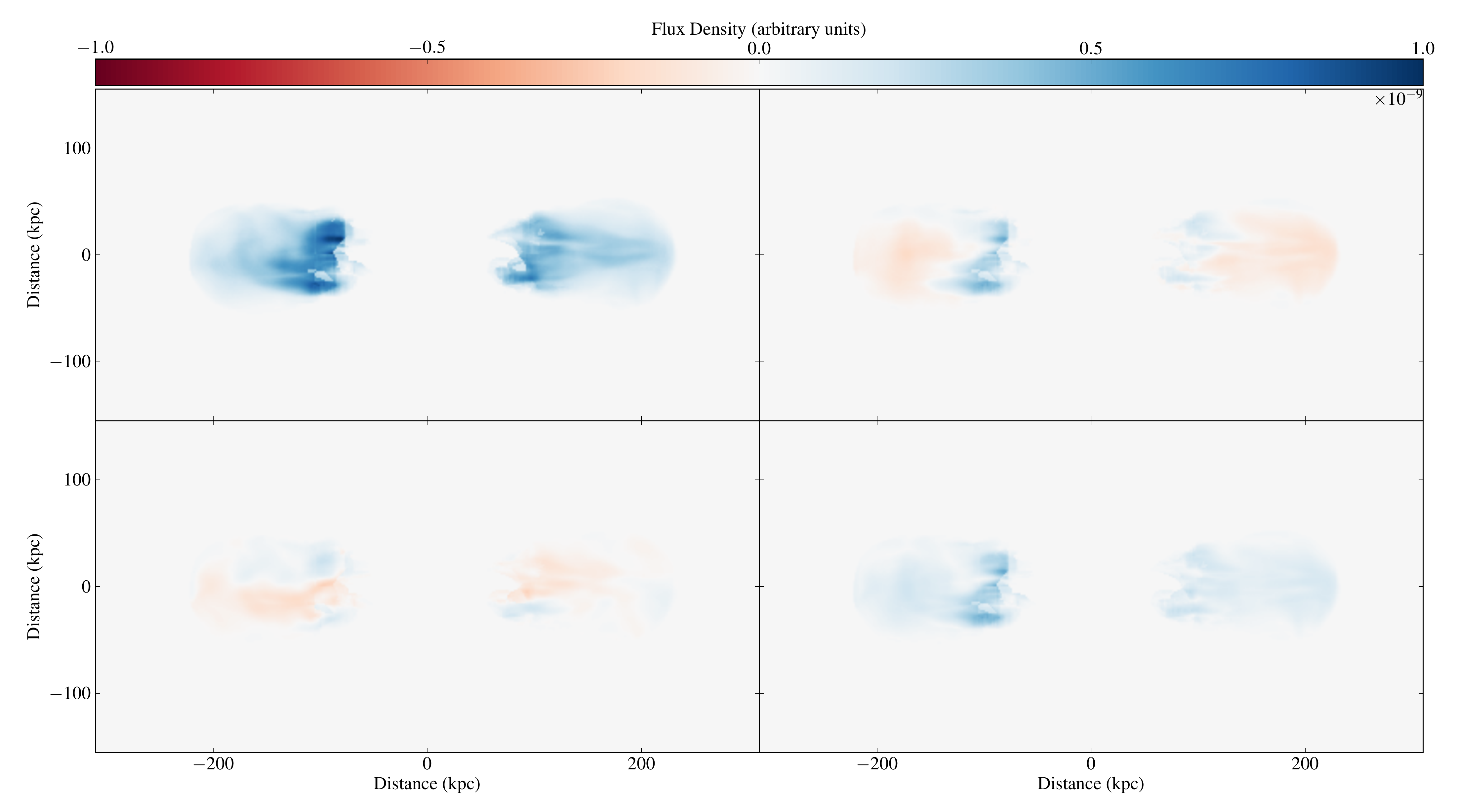}
\caption{Synchrotron emission maps for the r75-60 simulation, observed at 90 degrees to the jet axis at an age of 92.5 Myr, towards the end of the simulation. Top row: Stokes $I$ (left) and $Q$ (right). Bottom row: Stokes $U$ (left) and $P=\sqrt{Q^2+U^2}$ (right). All maps are scaled by the same arbitrary amount as in Figure~\ref{fig:SynchEarly} so that faint structure can be seen in all of the maps.}
\label{fig:SynchLate}
\end{figure*}

Figures~\ref{fig:SynchEarly} and~\ref{fig:SynchLate} show the decline
in surface brightness for our r75-60 model in a more visual form, with
Figure~\ref{fig:SynchEarly} showing the synchrotron emission maps from
a time-step shortly after the jet is switched off and
Figure~\ref{fig:SynchLate} showing the same model, with the same
scaling, at the end of the simulation. We can clearly see the drop in
emission across the entire source in the total synchrotron emission
maps (top left). Compact structures in both total and polarized
intensity have disappeared in the remnant source of
Figure~\ref{fig:SynchLate} and the surface brightness is much more
uniform than it was when the source was active. Observers should in
principle be able to use these features to help to decide whether a source is
a true remnant.

\section{Conclusions}

We have performed 3D RMHD numerical simulations of radio galaxies in
which the jets are shut off, in order to model the remnant phase and
test how the observable properties of the remnant depend on the
properties of the local cluster environment. In this paper the jets
  span a narrow range in jet power, one appropriate for powerful FRII
  radio galaxies, but probe a variety of environments through our
  choices of the cluster parameters $\beta$ and $r_c$. We reproduce
the same dynamical growth dependence on environment as seen in
\citetalias{HK13} and \citetalias{HK14} during the active phase, but
see that the environments' properties have little effect on the
evolution during the remnant phase, with all of the models proceeding
to grow at roughly the same rate. In the remnant phase, we find
  that the fraction of the input energy stored in the shocked shell
  increases rapidly with time, while the lobes continue to expand out
  of the cluster centre at speeds significantly higher than the
  buoyancy speed in most cases. The broad dynamics and energetics of these simulations are
  similar to those of others in which a jet in a realistic environment
is allowed to terminate
\citep{BA03,Zanni+05,Heath+07,Perucho+11,Ito+15,Hillel+Soker16,Chen+19}
though unlike other authors we focus on the phase immediately after termination.

We find that when the jets are no longer active the cluster gas begins
to settle back into the initial potential's distribution and the
weight of this infalling gas acts to push some of the dense cluster
gas along the jet axis, possibly along channels left by the jet. This
dense gas has two major effects on the remnant lobes. Firstly, it acts
to push the lobes further out of the cluster, at a faster rate than
simply buoyantly rising. As the lobes reach the lower density and
pressure regions of the cluster they undergo faster adiabatic losses,
and therefore cool faster than would be predicted by models in which
the unpowered lobes rise buoyantly out of the cluster. The second
effect of this dense cluster gas is to disrupt the structure of the
lobes. As the gas gets pushed along the jet axis, it impacts the rear
of the lobes and begins hollowing them out; eventually, in some
models, particularly those with high $\beta$ values, it creates a channel of cluster gas right through to the front
of the lobe.
                    
The evolution of surface brightness found in these models has a fairly
strong dependence on environmental properties. Both a smaller core radius and a
steeper density gradient lead to a significantly shorter time before
some threshold in surface brightness is reached, which we attribute to
the lobes being embedded in lower pressure cluster material at the
time the jets are shut off, resulting in a faster period of adiabatic
expansion. Thus, the remnant fraction predicted by simulations is
strongly dependent on the environment and the timing of the cessation
of jet activity, though in general we expect it to be shorter than it
is in analytic models, which do not capture all of the hydrodynamics.

Our models could be improved upon by attempting to solve the issue
of the internal boundary coupling well at early times by using models
with a higher resolution across the jet. We are limited here by
computational power and the two obvious ways around this have their
own drawbacks. By going back to 2D models we could achieve much
greater numerical resolution, but would lose the ability to create
realistic synthetic observations. Alternatively, we could make use of
\textsc{PLUTO}'s adaptive mesh refinement capabilities: however, this
is known to affect the small-scale magnetic field structure. Progress
is thus likely to require larger volumes in the simulations and more
computing power.

Another improvement would be to include a more realistic cooling
process, such as the methods of \citet{JRE99}, where a population of
electrons is carried around the grid like tracer quantities in a
number of momentum/energy bins. The radiative losses for this population
of electrons are then modelled at each time-step and updated, and the
resulting population can be
used to create much more realistic synthetic observations. Code to
implement these models is now under development and will be described
in a forthcoming paper.

\section*{Acknowledgments}

We are grateful to the referee, Manel Perucho, for constructive
  comments that have allowed us to improve the paper.

This work has made use of the University of Hertfordshire
high-performance computing facility
(\url{https://uhhpc.herts.ac.uk/}). WE thanks the UK Science and
Technology Facilities Council (STFC) for a studentship [ST/M503514/1].
MJH acknowledges support from STFC through grants [ST/M001008/1] and
[ST/R000905/1]. MGHK thanks the University of Tasmania for a Visiting
Fellowship.






\bsp	
\label{lastpage}
\end{document}